\documentclass[12pt]{emulateapj}
\usepackage{longtable}
\begin{document}
\title{A complex environment around Cir X-1}
\author{A. D'A\'i, R. Iaria, 
  T. Di Salvo,
  G. Lavagetto, N. R. Robba} 
\affil{Dipartimento di Scienze Fisiche ed Astronomiche,
  Universit\`a di Palermo, via Archirafi 36 - 90123 Palermo, Italy}
\email{dai@fisica.unipa.it}
\begin{abstract}
  We present  the results of  an archival 54 ks  long \textit{Chandra}
  observation of the peculiar source Cir X--1 during the phase passage
  0.223-0.261, based on  the phase zero passage at  the periastron, of
  its orbital period.  We focus  on the study of detected emission and
  absorption  features  using  the  High Energy  Transmission  Grating
  Spectrometer  on   board  of  the   \textit{Chandra}  satellite.   A
  comparative analysis  of X-ray  spectra, selected at  different flux
  levels of the  source, allows us to distinguish  between a very hard
  state, at a  low countrate, and a brighter,  softer, highly absorbed
  spectrum during  episodes of  flaring activity, when  the unabsorbed
  source luminosity is about three  times the value in the hard state.
  The  spectrum of  the hard  state  clearly shows  emission lines  of
  highly  ionized  elements,  while,  during the  flaring  state,  the
  spectrum  also shows  strong  resonant absorption  lines.  The  most
  intense and interesting  feature in this latter state  is present in
  the  Fe  K$\alpha$  region:  a  very broadened  absorption  line  at
  energies $\sim$ 6.5 keV that could result from a smeared blending of
  resonant   absorption  lines   of  moderately   ionized   iron  ions
  (\ion{Fe}{20}  -  \ion{Fe}{24}).  We  also  observe strong  resonant
  absorption lines  of \ion{Fe}{25} and \ion{Fe}{26},  together with a
  smeared absorption  edge above  7 keV.  We  argue that  the emitting
  region during  the quiescent/hard state  is constituted of  a purely
  photo-ionized medium,  possibly present above an  accretion disk, or
  of a photo-ionized  plasma present in a beamed  outflow.  During the
  flaring  states the  source undergoes  enhanced  turbulent accretion
  that modifies both  the accretion geometry and the  optical depth of
  the gas  surrounding the primary  X-ray source.  The  spectrum shows
  also  an  unidentified  broad  absorption  line at  $\sim$  3  sigma
  detection  level,  which  we  interpreted  as  a  highly  redshifted
  \ion{Fe}{26}  absorption  line.  Future observations  are,  however,
  needed to assess its real existence.
\end{abstract}
\keywords{line: identification -- line: formation -- stars: individual
  (Circinus X--1)  --- X-rays: binaries  --- X-rays: general}

\section{Introduction}
Cir X--1 is a peculiar  low-mass X-ray binary (LMXB), probably hosting
a neutron  star (NS), because of  the presence of  type-I X-ray bursts
detected  during an EXOSAT  observation \citep{tennant86}.   The first
long  term X-ray  lightcurve showed  a  periodicity of  $\sim$ 16.6  d
\citep{kaluzienski76}, that was interpreted as an effect of an orbital
period.  This   modulation  is  also   observed  in  the   radio  band
\citep{haynes78}, and in the IR \citep{glass94, clark03}.\\
Based  on  the presence  of  a  nearby  ($\sim$ 25  arcmin)  supernova
remnant, SNR G321.9-0.3, \citet{clark75} first hypothesized that Cir X-1
could  be a  runaway  binary system,  expelled  from the  site of  the
supernova explosion with a  high velocity kick.  The hypothesis gained
significant attention, also thanks to resolved radio maps, that showed
Cir X-1 is  embedded inside a radio nebula  \citep{haynes86} with jets
\citep{fender98}, bending toward the position of the SNR.\\
These jets have ultrarelativistic velocities \citep{fender04}, similar
to those produced by active galactic nuclei and galactic microquasars,
thus ruling  out the  belief that  only black holes  could be  able to
launch relativistic outflows.  The angle between our line of sight and
the direction  of the  jet was inferred  to be less  than 5$^{\circ}$.
More  recently,  the  jets  have  been also  resolved  in  the  X-rays
\citep{heinz07}.\\
The connection between the SNR  G321.9-0.3 and Cir X-1 seems, however,
now  to  be  compromised   by  some  \textit{Hubble  Space  Telescope}
observations \citep{mignani02}, that  showed no apparent proper motion
(with an upper limit of 200  km/s), contrary to what had been expected
\citep{tauris99}.\\
The  16.6 cycle,  although  first  detected in  the  X-rays, has  been
largely studied in  the radio wavelength. Based on  the onset times of
radio flares observed between  1978 and 1988 \citet{stewart91} derived
the  ephemeris of  the  source,  giving, for  phase  0, the  following
formula:
\begin{equation}
MJD_{N} = 43076.37 + N (16.5768 - 3.53 \times 10^{-5}N),
\end{equation}
that constituted, for more than  a decade, the reference ephemeris for
the source.  Recently \citet{saz03},  using only X-ray data spanning a
period of 30  years, obtained the first X-ray  ephemeris. However, the
analysis was based on the occurrence  of periods of X-ray maxima in the
X-ray lightcurves,  that \citet{clarkson04} argued could  be flawed by
large intrinsic  phase scattering, thus leading  to systematic errors.
Adoption of the periastron X-ray dips was found to provide, therefore,
a more suitable system clock, thus resulting in the new ephemeris
\begin{equation}
MJD_{N} = 50081.76 + N (16.5732 - 2.15 \times 10^{-4}N).
\end{equation}
This  solution implies,  moreover, a  significant high  orbital period
derivative,  resulting  in  a  characteristic  life  time-scale,  $P/2
\dot{P}$,  for the  system of  $\sim$ 1000  years.  \citet{clarkson04}
concludes  that Cir X-1  is, indeed,  a very  young system,  whose SNR
could   be  the   same  radio   nebula  in   which  it   is  embedded.
\citet{tudose06} derived,  under the assumption that  the radio nebula
is, instead, powered by the jet activity, a nebula life time in the $4
\times 10^3 - 4 \times 10^5$ yr range and a jet injected power of
$\sim 10^{35}$ erg s$^{-1}$.\\
The companion star in the Cir X-1 has been long searched, but the high
visual extinction  towards the direction  of the source  has prevented
its determination.  Recently, \citet{jonker07}, through optical I-band
monitoring  of  the  emission/absorption  Paschen lines  seen  in  the
optical spectra of the system and assuming that the lines are produced
by  the   companion  star,  have  derived   the  best-fitting  orbital
parameters  of the  system: an  eccentricity  $e$ =  0.45, an  orbital
period =  16.6 d  and a$sin(i)$ =  16.9 light-seconds.   The companion
star is, assuming a mass for  the compact object of 1.4 M$_{\odot}$, a
supergiant  of B5-A0  stellar type,  whose  mass should  be $\leq$  10
M$_{\odot}$ while the inclination of
the system should be $\geq$ 14$^{\circ}$. \\
This X-ray source shows some spectral behaviors similar to those shown
by black-hole (BH) binaries, but others that are most similar to those
shown  by  NS  binaries,  making  the  nature  of  this  system  quite
enigmatic.   The  X-ray spectrum  of  Cir  X--1  has been  extensively
studied  in  the past  for  each phase  of  the  orbital period.   The
continuum   emission  consists  of   a  thick   Comptonized  component
\citep{iaria01a}, which,  for relatively narrow  band coverage, cannot
be distinguished from  a simpler sum of two  thermal components, and a
variable hard-tail above 10  keV \citep{iaria01b}.  The spectrum shows
a  phase-dependent  evolution, whose  major  characteristic, when  the
source is  near the  periastron passage is  the appearance of  a large
column  density  of  neutral  matter  occulting  the  continuum  X-ray
emission. Its origin  has been assigned or in  the colder outer layers
of the accretion  disc, or alternatively in the  occulting presence of
the  companion  star   \citep{brandt96}.   The  first  interpretation,
however, implies an edge-on geometry which is in contrast with the
observed small angle of view of the jet.\\
High-resolution spectra obtained with \textit{Chandra} \citep[][hereafter Paper
I]{schulz02} showed  P Cygni  profiles of H-  and He-like  ions, whose
intensity decreased  with decreasing luminosity of the  source.  The P
Cygni structures  were all blue-shifted  with respect to  the expected
laboratory-frame energies,  thus leading to the  conclusion that these
features could arise from a radiatively, or thermally-driven wind from
a  hard X-ray  irradiated  accretion disc.   The  spectrum also  shows
complex absorption features above 7 keV, whose intensity and position
vary according to the orbital phase \citep{iaria01a}.\\
Timing properties have also been also extensively studied in the past,
but recently  \citet{boutloukos06} has shown, using a  large sample of
archival RXTE  data, that the timing  behavior is very  similar to the
one displayed  by the  bright accreting NS  systems, belonging  to the
class  of the  Z sources.   The mostly  notable characteristic  is the
presence  of twin kilohertz  quasi-periodic oscillations  (kHzQPOs) in
the power spectrum of the source at peak frequencies in the 56--223 Hz
range,  for  the lower  peak,  and 230--550  Hz  for  the upper  peak.
However,  the   frequency  range  of   the  kHzQPOs  in  Cir   X-1  is
significantly lower than the one shown by other accreting NS sources
\citep[see also][]{belloni07}.\\
Cir X-1 seems, therefore, to  share the basic ingredients of the class
of  the old,  low magnetized,  bright  NSs system,  although the  high
inferred eccentricity,  the variable phase  dependent spectral changes
and  the very  probable young  age  do not  fit into  the classical  Z
classification scheme.   Observations of spectral  and timing features
common to physical objects  of such different physical characteristic,
therefore, offer a unique opportunity to test present theories against
experimental  data. \\  In this  work  we  present  a detailed  spectral
analysis  of Cir  X-1  using a  54  ks long  \textit{Chandra} observation.   We
analyzed the  source when  it was highly  variable, but only  within a
restricted  orbital phase.   This allowed  us to  investigate  how the
emission and absorption lines change with the continuum without having
to consider the complicating effects  of the change in phase.  Through
X-ray spectra of this source selected at different flux levels we show
the  complex spectral changing  of the  source; data  clearly indicate
that the emission line features  are present both during the quiescent
state and the flaring state of the source, while it is only during the
flaring activity,  when the source countrate strongly  rises, that the
spectrum shows  strong absorption features of highly  ionized iron and
calcium. The appearance  of these features is linked  to the continuum
emission which  strongly varies, becoming,  from the quiescent  to the
flaring  states,  more luminous  but,  at  the  same time,  completely
obscured by a large column density of cold material.
\section{Observation and data reduction} 
\begin{figure*}
\centering
  \includegraphics[height=8cm, width=6cm,  angle=-90]{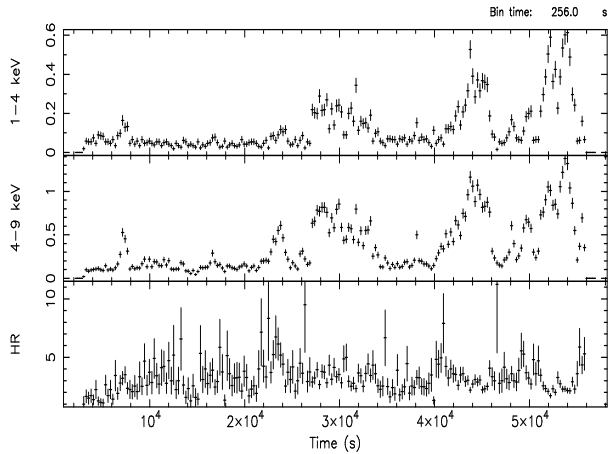}
  \caption{Lightcurves of Cir X-1 \textit{Chandra} observation.
Top panel: a 1.0--4.0 keV lightcurve of Cir X-1 \textit{Chandra} 
observation extracted from selected regions around the HEG and MEG first order arms.
Center panel: 4.0--9.0 keV lightcurve using the same regions of top panel.
Bottom panel: hardness ratio of the energy bands [4.0--9.0]/[1.0--4.0].}
\label{fig1}
\end{figure*}

\begin{figure*}
\centering
\includegraphics[height=8cm, width=6cm,  angle=-90]{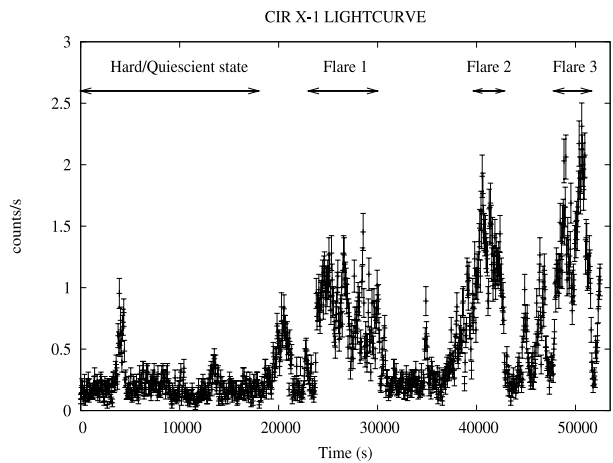}
\caption{Lightcurve of Cir X-1 \textit{Chandra} 
observation extracted from selected regions around the HEG and MEG first order arms.
Above the lightcurve, arrows delimit  the time-intervals that we used
to extract the time-selected spectra.
Bin-time in lightcurve is set to 64 s.}
\label{fig2}
\end{figure*}

Cir X--1  was observed  with the \textit{Chandra}  observatory on 2005  June 02
from   08:35:58  to   23:44:51  UT   using  the   \textit{High  Energy
  Transmission Grating Spectrometer}  (HETGS), for a total integration
time of  52650 s (OBS.  ID.   5478, from the  \textit{Chandra} public archive).
The   corresponding   orbital    phase,   using   the   ephemeris   of
\citet{stewart93}, lies  between 0.065 and  0.104, based on  the phase
zero   passage  at   the  periastron.    Adopting  the   ephemeris  of
\citet{clarkson04}, this  corresponds to the  phase range 0.223-0.261.
The observation was performed in timed faint mode, adopting a subarray
(512 rows)  of the ACIS-S detector  to mitigate the  effects of photon
pile-up, with a CCD frame time  of 1.7 s.  We processed the event list
using available  software (FTOOLS v6.1.2 and CIAO  v3.2 packages).  We
computed aspect-corrected exposure maps for each spectrum, allowing us
to  correct   for  effects  from   the  effective  area  of   the  CCD
spectrometer.  The brightness of the source prevented the study of the
zeroth-order events since these are  mostly affected by pile-up On the
other hand, the grating spectra are not, or only moderately (less than
10 \%), affected.   In this work we utilize the  1st-order HEG and MEG
spectra.  Data were extracted from regions around the grating arms; to
avoid overlapping between  HEG and MEG data, we used  a region size of
25  and  33  pixels for  the  HEG  and  MEG, respectively,  along  the
cross-dispersion direction.  The  background spectra were computed, as
usual, by extracting data above and below the dispersed flux.  We used
the standard CIAO tools to  create detector response files for the HEG
-1 (MEG -1) and HEG +1 (MEG +1) order (background-subtracted) spectra.
To compute as  accurately as possible the position  of the zero order,
we extracted  the event  list during a  time interval when  the source
showed the lowest countrate and variability; from the zero-order image
we derived the counts/pixel grid.  The center of the pixel showing the
highest counting corresponds to  the FK5 coordinates R.A.  $15^h :20^m
:40.87^s$  and Dec.   $-57^h :10^m  :00.24^s$,  with an  error on  the
coordinates of 0.6$^{''}$. The identification  of the zero order is in
agreement with  the result of \citet{iaria07}, when  the source during
another \textit{Chandra} observation was in a very low luminosity state and the
zero order was  not affected by pile-up issues.   After verifying that
the negative  and positive orders  were compatible with each  other in
the  whole  energy  range  we  coadded  them  using  the  script  {\it
  add\_grating\_spectra} in the CIAO software, obtaining the 1st-order
MEG spectrum and the 1st-order HEG spectrum.  HEG and MEG spectra were
finally  rebinned in  order  to have  at  least 25  counts per  energy
channel to allow  the use of $\chi^2$ statistics. We  used MEG and HEG
spectra in the 1.3--5.0 keV and  in the 1.5--9.0 keV energy range.  In
Figure~\ref{fig1} we  show the 256  s bin-time lightcurve  taking into
account only  the events  in the first-order  HEG and  MEG diffracting
regions in two energy bands.   Upper panel shows the temporal behavior
of the low energy photons (1.0--4.0 keV), while center panel shows the
same  in the  high energy  band (4.0--9.0  keV).  The  two lightcurves
follow the  same pattern of variability,  with a low  countrate at the
beginning  of the  observation, lasting  less than  20 ks,  with three
short-term peaks; the second  part of the observation is characterized
by intense source variability with three broad peak profiles where the
countrate rises by more than an  order of magnitude in both bands.  In
the bottom  panel of the curve we  show the hardness ratio  of the two
energy  bands which  is slightly  anticorrelated with  respect  to the
countrates  \citep[see   also][~for  a  discussion   on  the  spectral
hardening evolution of the source]{shirey99}.\\ The lightcurve for the
whole energy  band, for  the same extracting  regions, has  an average
countrate   of   $\sim$   0.47   counts/s   and   it   is   shown   in
Figure~\ref{fig2}.\\  Given the  strong variability  of the  source we
performed  a time-selected  spectral analysis  based on  the countrate
variability of the source.  The  first time-selected state lies in the
first  part of  the  observation,  when the  source  shows the  lowest
countrate and  presents the minimum  level of variability.   From this
time selection we, however, excluded  two short minor peaks present in
the lightcurve.   The total collecting  time for this state  is $\sim$
17.3  ks,  with  an  average  countrate of  0.35  cts/s.   The  second
time-selected spectrum encompasses the  peak of the first broad flare,
for a time duration of 7.3 ks with an average countrate of 1.38 cts/s.
The third  state is centered  around the second, sharper,  peak, while
the fourth state  is centered in the last  segment of the observation.
The collecting time and the average countrate for these selections are
3.10 ks  and 1.15 counts/s,  3.84 ks and 1.32  counts/s, respectively.
We show in Figure~\ref{fig2} the lightcurves of the entire observation
and  of the  time-selected intervals.\\  We considered  negligible the
level  of  pile-up  for  our  observation.  We  left  a  normalization
constant, between the  HEG and MEG spectrum free to  vary, in order to
take into  account residual  flux calibration uncertainties.   For the
spectral analysis we used the software package Xspec 11.3.2p.
%%%%%%%%%%%%%%%%%%%%%%%%%%%%%%%%%%%%%%%%%%%%
%%%%%%%%%%%%%%%%%%%%%%%%%%%%%%%%%%%%%%%%%%%

\section{Spectral Analysis}
\subsection{The time-selected spectra: the quiescent state}
In order to  fit to the continuum emission of the  source we adopted a
spectral   decomposition   that   closely  follows   previous   models
\citep[e.g.~][]{brandt96,iaria01a,  schulz02}:   an  absorbed  thermal
component  ({\it bbody}  in Xspec)  multiplied by  a  partial covering
component ({\it pcfabs} in Xspec), while the value of the interstellar
absorption ({\it  wabs} in Xspec) was  frozen to a  reference value of
1.85  $\times$ 10$^{22}$  cm$^{-2}$ adopting  the  cross-sections from
\citet{morrison83}.   The  partial   covering  component  is  strictly
required  by  the fit;  a  simpler {\it{wabs  *  bb}}  model gives  an
unrealistically  high value  for the  blackbody  temperature (kT$_{BB}
\geq  9$ keV)  and a  corresponding  high value  for the  interstellar
equivalent  hydrogen column.   The blackbody  temperature is  $\sim$ 3
keV, with a  corresponding blackbody radius of 0.5  $\pm$ 0.1 km.  The
model is  insensitive to the addition of  further spectral components.
The  blackbody   component  can  be  replaced  by   a  flat  power-law
(photon-index =  -0.1$_{-0.5}^{+0.4}$), without significant  change in
the $\chi^2$  value of the fit  and in the  other spectral parameters.
The continuum  flux is significantly obscured by  the partial covering
effect,  with  an  associated   equivalent  hydrogen  column  N$_H$  =
16$^{+5}_{-3}$ $\times$  10$^{22}$ cm$^{-2}$, and  a covering fraction
$f$  = 0.69$^{+0.07}_{-0.06}$.   The shape  of the  spectrum  is quite
peculiar, and, as far as we  know from literature, this is the hardest
spectrum  ever observed  in this  source.   The absorbed  flux in  the
\textit{Chandra}  energy  range  (1.0--10  keV) corresponds  to  1.56  $\times$
10$^{-10}$  erg cm$^{-2}$  s$^{-1}$,  while extrapolating  in a  wider
range (0.01--100  keV) results in  a flux of 4.63  $\times$ 10$^{-10}$
erg  cm$^{-2}$  s$^{-1}$.   We  note, however,  that  this  simplified
continuum model, although it clearly indicates a rather hard spectrum,
is not suited  for the study of the continuum  behavior of the source,
given  the restricted  energy  band  of \textit{Chandra},  and  a more  complex
spectral decomposition is expected for  this kind of source.  In fact,
the  continuum emission,  that  we fitted  with  a single  temperature
blackbody, does  not reproduce a  typical accreting NS  spectrum, that
has softer  thermal temperatures and  present generally two,  or more,
temperature  peaks.  The  following analysis  will,  therefore, mainly
focused  on  the  detection  of  narrow  discrete  absorption/emission
features.  We  tested their  presence by adopting  different continuum
models (as the best-fitting  model discussed above, polynomial fits in
a narrow  band around the  features, power-law fits), and  although we
observed slight  changes in their detection  thresholds, or parameters
space, we found  that the overall picture and  the conclusions derived
are  not affected  by the  particular continuum  that was  adopted. We
caution, however, the reader that a significant discrepancy, dependent
on the model  adopted, can arise for the  determination of the optical
depth of iron edges above 7 keV, as the continuum cannot be sufficiently
well constrained above these energies.
The  spectrum  is  characterized  by  some  clear  discrete  features;
adopting our best-fit continuum, we detect, with a significance always
above  3 $\sigma$, three  emission lines  at energies  consistent with
resonant emission  lines of  H-like Si, He-like  Ca and He-like  Fe, a
rather narrow  absorption line at 6.49  $\pm$ 0.03 keV, and  a 6.4 keV
emission  line, possibly  associated with  fluorescence  emission from
neutral,  or lowly  ionized,  iron.  Above  7  keV we  found a  strong
absorption edge at energies  correspondent to the neutral Fe K$\alpha$
edge.  The emission lines from  highly ionized elements are narrow, as
we find only  upper values to the line  widths (7 eV, 25 eV  and 28 eV
for  \ion{Si}{15}, \ion{Ca}{19}  and \ion{Fe}{25}  respectively).  The
detection of  the iron  fluorescence line at  energies $\sim$  6.4 keV
indicates  the presence  of a  cold reflecting  medium, which  is also
responsible for the sharp edge  observed at higher energies.  From the
optical  depth of the  edge ($\tau$  = 1.2$^{+0.3}_{-0.4}$),  we infer
that the associated equivalent hydrogen column is (6 $\pm$ 2) $\times$
10$^{24}$ cm$^{-2}$,  a very  high value that  exceeds by an  order of
magnitude the  value reported by \citet{shirey99} when  the source was
in a dip-state.   We also tested the possibility  that this edge could
be  partially  smeared using  the  {\it  smedge}  component in  Xspec,
\citep{ebisawa91},  keeping  fixed  the  index for  the  photoelectric
cross-section at the default value of  -2.67. We did not find, in this
case,  any evidence  of  smearing, with  only  an upper  value of  the
smearing width of  0.2 keV.  We report in  Table~\ref{tab1} the values
and  the  uncertainties  for   the  best-fitting  parameters  for  the
continuum   emission,  while   in  Table~\ref{tab2}   we   report  the
corresponding  parameters for  the detected  lines. Errors  are always
reported  at $\Delta  \chi^2 =  2.7$. We  show in  the left  panels of
Figure~\ref{fig3},  Figure~\ref{fig4}, Figure~\ref{fig5}  the  plot of
data,  with  superimposed the  best-fit  model  in  the energy  ranges
1.4--2.7 keV, 3.8--4.2 keV and 6.2--7.4 keV.

\subsection{The time-selected spectra: the flaring states}

During  this  \textit{Chandra} observation  we  observe  the source  undergoing
episodes  of flaring activity;  we distinguish  three long  flares, of
approximatively 8  ks, 5.4  ks and  4 ks duration,  and 7  short minor
flares.   We time selected  three energy  spectra, extracted  from the
lightcurve, centered  at the  countrate peaks of  the long  flares and
performed a  spectral analysis using the same  data reduction criteria
of  the quiescent  state extraction.   We will  simply refer  to these
spectra as Flare 1, 2 and 3 spectrum for the first, the second and the
third  long  flare  present   in  the  lightcurve,  respectively  (see
Figure~\ref{fig2}).   For  the continuum  emission  of  the source  we
adopted the same model of  the quiescent state, with the same spectral
parameters free to vary.\\ The  spectra show in all the flaring states
a significant change both in the primary continuum emission and in the
absorbing  medium  with  respect  to  the hard  state.  The  blackbody
temperature drops to $\sim$ 1.6  keV from the $\sim$ 3 keV temperature
of the  quiescent state, while the unabsorbed  luminosity results more
than  doubled  (see  Table~\ref{tab3}).   We  note,  however,  that  a
power-law  component  could replace  the  blackbody component  without
significantly changing the $\chi^2$ value  of the fit. In this case we
found for all the flaring spectra  value of the power-law index in the
1.5--2.2 range.   The partial covering  matter during the  first flare
has the same equivalent hydrogen  column as during the quiescent state
while  the covering  fraction of  the  obscuring medium  covers up  to
$\sim$  94 \%  of the  total  continuum emitting  region.  During  the
second  and the  third flare,  the continuum  flux continues  to rise,
while the covering fraction of the obscuring medium is close to unity,
and the  equivalent hydrogen column progressively  reduces.  The flare
selected spectra  are, thereafter, much  softer but at the  same time,
highly photoelectrically  absorbed, so that we have  an extremely poor
statistics below  2 keV.  The absorbed  fluxes in the  1--10 keV range
are 5.6 $\times$ 10$^{-10}$, 7.6 $\times$ 10$^{-10}$, 8.5
$\times$ 10$^{-10}$ erg cm$^{-2}$ s$^{-1}$, for Flare 1, 2 and 3 respectively.\\

The  emission evolution  of the  continuum  is also  accompanied by  a
substantial modification of the  spectral features of the source, that
is more remarkable  in the 6.2--7.4 keV iron  region.  All the flaring
spectra do not present apparent  emission lines, but, on the contrary,
we observe mostly absorption resonant lines of H-like and He-like Ca
and Fe, together with a broadened  Gaussian in the  6.5--6.6 keV energy range.\\
The  H-like resonant  iron  absorption  line is  detected  in all  the
flaring  spectra,  as  well  as   the  H-like  calcium  line  (with  a
significance of $\sim$ 4 $\sigma$ for each spectrum); the widths of
both  lines  are  not  well  constrained  by  the  fit,  although  the
best-fitting  values  of  the  \ion{Ca}{20}  are  consistent  with  an
intrinsic narrowness, while the \ion{Fe}{26} line, overall in state 1,
is broadened.\\  The He-like calcium absorption line  is well detected
in two of the  three flaring spectra; in Flare 1 we  had to freeze the
value of the line energy  to the expected rest-frame value, given that
the line appeared rather broadened,  while in Flare 3 the derived line
energy  is consistent  with the  expected value.  The He-like  Ca line
appears,  contrary to what  found for  the corresponding  H-like line,
intrinsically broadened and more intense.\\

\begin{figure*}
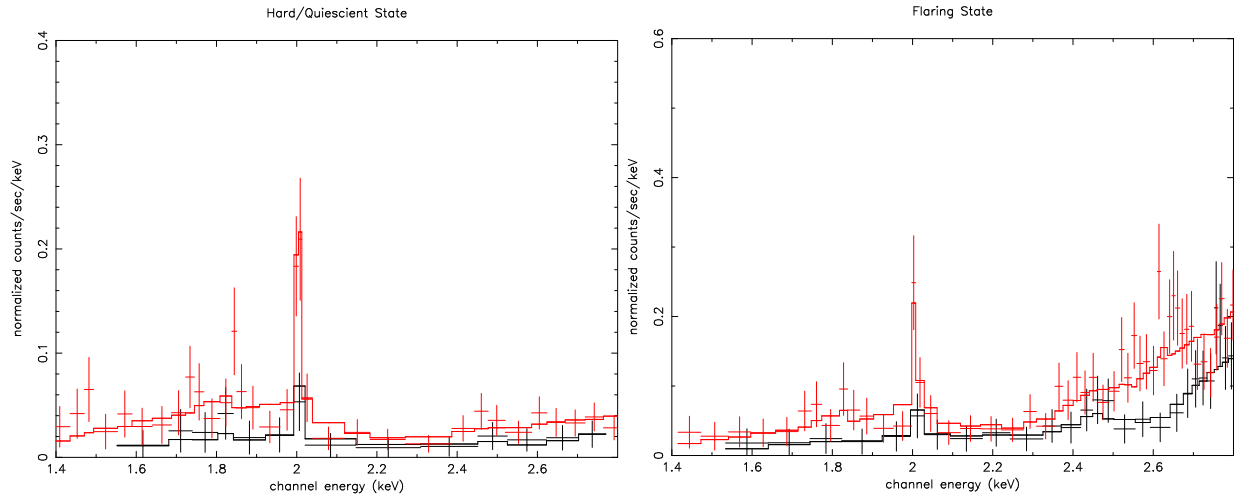

\centering
\includegraphics[angle=-90, scale=0.35]{f3a.eps}
\includegraphics[angle=-90, scale=0.35]{f3b.eps}
\caption{Left  panel:  data  together   with  best-fit  model  in  the
  low-energy band  (1.4-2.7 keV) during the hard/quiescient state.
  Right panel:  same region in the total flare time integrated spectrum. 
  The only feature clearly seen in both spectra is the Ly$\alpha$ emission line 
  of \ion{Si}{14} at $\sim$ 2 keV. MEG data are shown in red, HEG data in black.}
\label{fig3}
\end{figure*}

\begin{figure*}
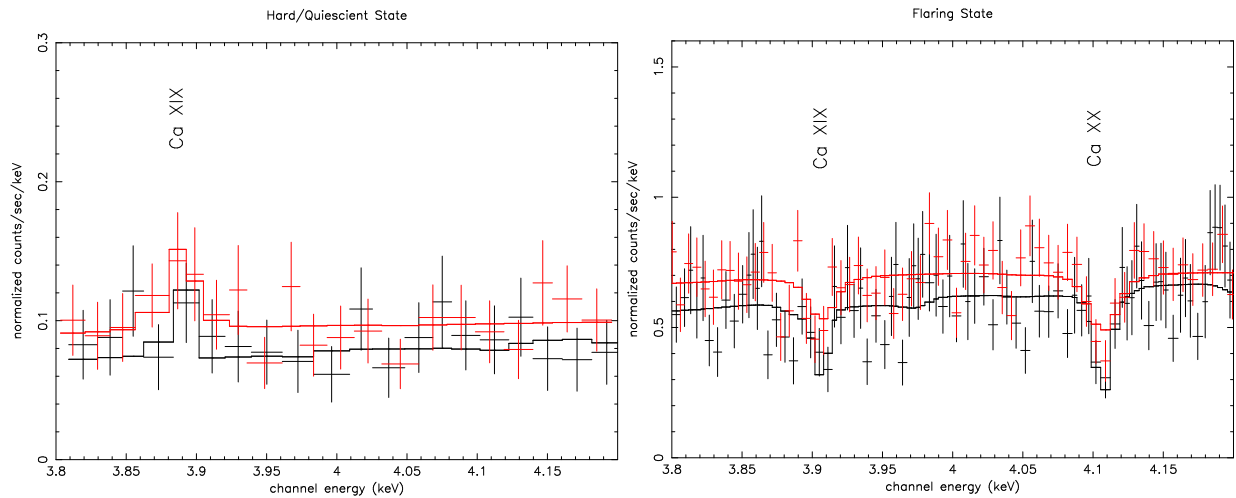

\centering
\includegraphics[angle=-90, scale=0.35]{f4a.eps}
\includegraphics[angle=-90, scale=0.35]{f4b.eps} 
\caption{Plot  of  data  with  best-fit  model  superimposed  for  the
  quiescent/hard state and  for the flaring state in  the 3.8--4.2 keV Ca
  region. Best-fitting  parameters and  associated  errors for  the
  continuum emission and of the line emission features are reported in
  Table~\ref{tab1} and in Table~\ref{tab2} respectively. MEG data are shown in red, HEG data in black.}
\label{fig4}
\end{figure*}

\begin{figure*}
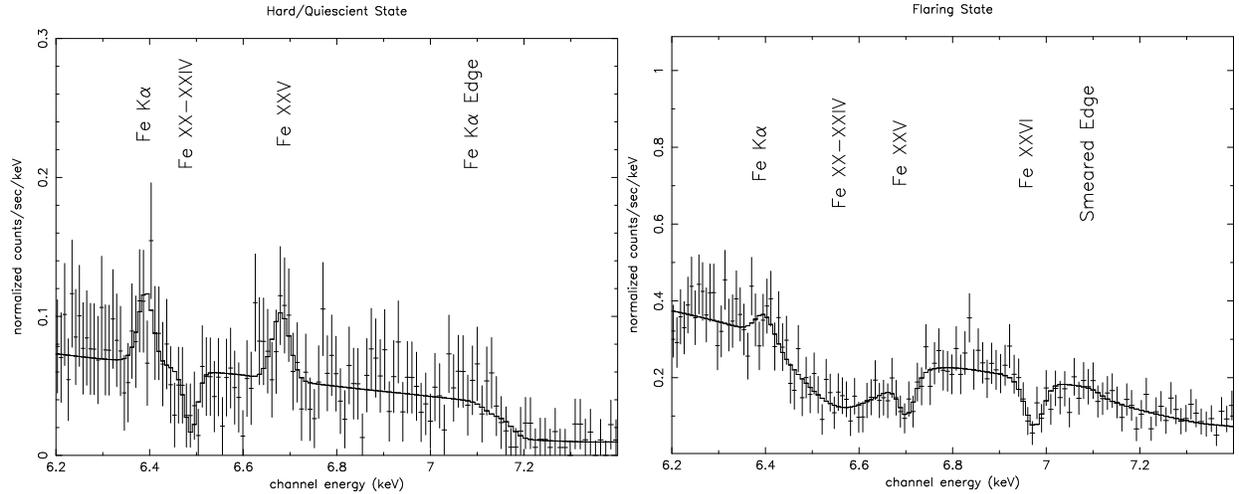

\centering
\includegraphics[angle=-90, scale=0.35]{f5a.ps}
\includegraphics[angle=-90, scale=0.35]{f5b.ps} 
\caption{Plot  of  data (no rebin applied) with  best-fit  model  superimposed  for  the
  quiescent/hard state and  for the flaring state in  the 
  6.2--7.4 keV  Fe region. Best-fitting  parameters and  associated  errors for  the
  continuum emission and of the line emission features are reported in
  Table~\ref{tab1} and in Table~\ref{tab2} respectively.}
\label{fig5}
\end{figure*}

The He-like  iron line  is detected  only in the  last flare,  but its
position could  not reasonably be constrained  by the fit  and we kept
frozen the line  energy to the laboratory rest frame.   We do not find
any evidence of possible Fe K$\alpha$ line; by fixing the position and
the  width of  the  line to  the best  values  of the  fit during  the
quiescent  state we  derived  from  the normalization  of  the line  a
significance of  only $\sim$ 1  $\sigma$ for all the  three flaring
spectra.  We observe in the Fe K$\alpha$ region, for the first time in
this  source, a complex  broadened feature  in absorption  at energies
between  6.54 and  6.64  keV that  can  be simply  fitted  by a  broad
Gaussian in absorption;  we find that this feature  is present in each
flaring  state, with the  energy of  the line  centroid that  does not
significantly vary during the three flaring episodes; the width of the
line is not always well constrained by the fits, particularly in state
2 where the relative error value is about 50 \% of the best-fit value;
however, for all  the states, we find a  common range, consistent with
all the  states, in  the 70--100  eV range.  Fits  with two,  or more,
Gaussians replacing the single Gaussian did not provide significant
changes in the $\chi^2$ value of the fit.\\
In each  flaring state we  find that spectrum  falls above 7  keV more
rapidly than the expected best-fit model. We, therefore, added an edge
component, as also expected to  be present from the complex absorption
features in  the 6.4--7.0 keV  energy range.  Using  the \textit{edge}
component we find, for each  flare state, a threshold energy at $\sim$
7.4  keV, with an  optical depth  that is  almost constant  during the
different  flares at $\tau  \sim$ 0.5.   Replacing the  edge component
with the smeared edge component \textit{smedge}, slightly improves the
$\chi^2$ value of the fit, although not significantly and, generally,
we find only upper values to the smearing width (see Table~\ref{tab3}). \\
Given  the   presence  of  the   line  features,  and   their  general
consistency,  in the three  flaring spectra  and the  smooth continuum
evolution of the source, we decided  to merge the three spectra into a
single  spectrum in  order to  increase the  statistics and  to better
constrain features present  in each flare state.  Although  there is a
significant change  in the continuum  parameters (mainly in  the total
luminosity of the source and in the N$_H$ value of the \textit{pcfabs}
component),   our   main   focus   is  addressed   to   the   detected
emission/absorption  features and we  will use  the continuum  for the
detection  of  the local  emission/absorption  lines.   We report  the
best-fit  continuum parameters  and  associated errors  in the  second
column of Table~\ref{tab1}, while  the parameters of detected features
are  summarized in  Table~\ref{tab2}.  The  merged  ``flare'' spectrum
shows emission  features that were  statistically hidden by  the short
exposure  times.  However  we choose  to fix  line energy  or/and line
width to  the value  of the total  time integrated spectrum  (see next
section) in order to derive a better constraint to the line fluxes, in
cases where the fit could not constraint at the same time all the line
parameters.  In  particular, the spectrum shows  the resonant emission
line   of   \ion{Si}{14}  and   \ion{S}{16}   (see   right  panel   of
Figure~\ref{fig3}) and the Fe K$\alpha$  emission line at 6.4 keV.  It
is  to be  noted,  however, that  the  flux of  the \ion{Si}{14}  line
results much higher with respect to the corresponding value during the
hard  state.  We  do not  find  any evidence  of the  presence of  the
\ion{Ca}{19}   emission  line,  but   we  observe   the  corresponding
absorption line,  together with the \ion{Ca}{20}  absorption line (see
right panel  of Figure~\ref{fig4}).  The  iron region is  dominated by
three  broad absorption  lines: the  blended \ion{Fe}{20}-\ion{Fe}{24}
line at 6.57 keV line, the  He-like Fe absorption line at 6.70 keV and
the H-like  line at 6.97  keV.  A plot  of the data  with superimposed
best-fit model for  this energy region is shown in  the right panel of
Figure~\ref{fig5}.
%%%%%%%%%%%%%%%%%%%%%%%%%%%%%%%%%%%%%%%%%%%%%%%%%
%%%%%%%%%%%%%%%%%%%%%%%%%%%%%%%%%%%%%%%%%%%%%%%%%
\subsection{The total integrated spectrum}
\begin{figure*}
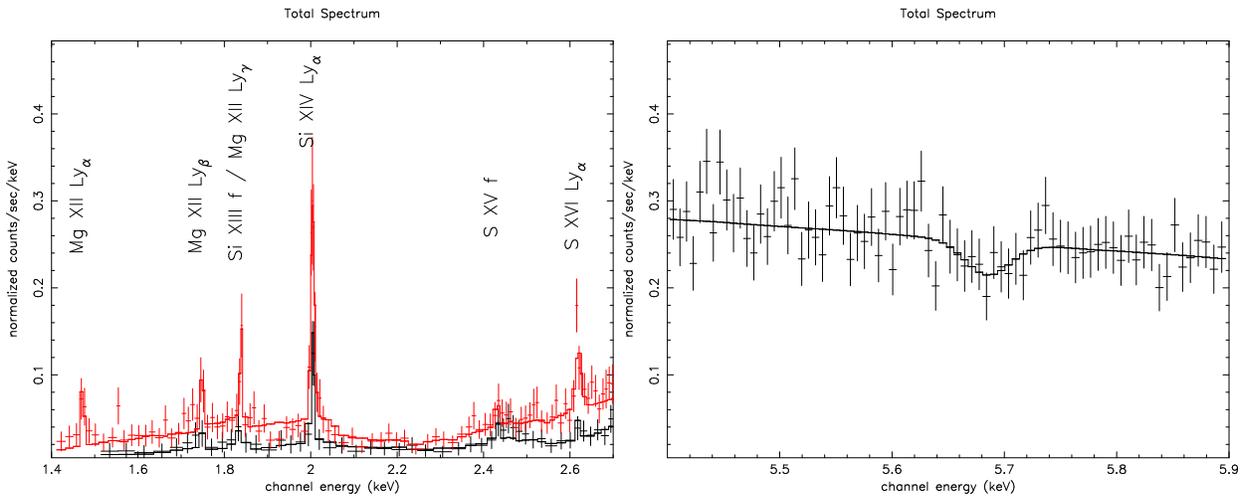

\centering
\includegraphics[angle=-90, scale=0.35]{f6a.eps}
\includegraphics[angle=-90, scale=0.35]{f6b.ps}
\caption{Left  panel:  data  together   with  best-fit  model  in  the
  low-energy band  (1.4-2.7 keV) for the total time integrated spectrum.  Right panel:
  an identified absorption feature in the same spectrum at 5.689 keV. See discussion.}
\label{fig6}
\end{figure*}

As a  final step in our  analysis we produced a  time integrated first
order HEG/MEG  spectrum of  the entire observation,  as it  will prove
helpful for  detecting features that  are statistically damped  by the
low  S/N of short-time  exposure spectra.   The continuum  spectrum is
obviously to be taken as rather  indicative of the merging of the hard
quiescent state  and of  the softer, but  much more  absorbed, flaring
state.  The best-fitting  values of the continuum, in  fact, reflect a
weighted average  of the changing spectrum  of the source,  and, as in
the case of the merged flare spectrum, this will mainly serve to put a
continuum  under the  detected  features.  For  completeness' sake  we
report, however, these values in the last column of Table~\ref{tab1}.\\
On  the soft  part of  the spectrum  the emission  lines  belonging to
H-like   ions  of   \ion{Mg}{12}   (Ly$_{\alpha}$,  Ly$_{\beta}$   and
Ly$_{\gamma}$),    \ion{Si}{14}   (Ly$_{\alpha}$)    and   \ion{S}{16}
(Ly$_{\alpha}$)  are clearly  detected.   However, the  high flux  and
equivalent  width of  the \ion{Mg}{12}  Ly$_{\gamma}$, which  are both
comparable  to  the  value   of  the  \ion{Mg}{12}  Ly$_{\beta}$,  are
indicative  of  a  possible   blending  with  the  forbidden  line  of
\ion{Si}{13}, whose  rest-frame energy position is at  1.8397 keV.  We,
then, looked  for the corresponding He-like resonance  lines for these
elements by  locally adding Gaussian  emission lines, fixing  the line
energies  at 1.3522,  1.8449  keV, and  2.4606  keV for  \ion{Mg}{11},
\ion{Si}{13}  and  \ion{S}{15},  respectively  and leaving  width  and
normalization of each line as free parameters.  The significance of
these lines resulted always less  than 2 $\sigma$.  Given the highly
probable presence of the forbidden line of \ion{Si}{13}, we looked for
the   other   forbidden    lines   of   \ion{S}{15}, \ion{Ca}{19}   and
\ion{Fe}{25}.   Fixing  the centroid  line  energies  to the  expected
rest-frame values  (2.4307 keV,  3.8612 keV and  6.6366 keV),  and the
widths of  the lines at 2 eV  (fit was not sensitive  to variations of
this  parameter) we obtained  a significance  of 2.6  $\sigma$, 2.8
$\sigma$  and  less than 1  $\sigma$  for the  \ion{S}{15},  \ion{Ca}{19}  and
\ion{Fe}{25}, while  the best-fit equivalent  widths were respectively
10 eV  3 eV and  4 eV. Although  the significance of  the forbidden
line of  \ion{Fe}{25} is  very low, we note  that the spectrum  is strongly
curved  by the  broad  absorption  feature at  $\sim$  6.5 keV,  that,
together  with  the  merged  changing  continuum, can  have  a  strong
smearing effect  on this emission  line. A closer inspection  of these
features, using  unbinned spectra,  revealed that the  centroid energy
lines  well  match  the  peak   of  the  Gaussian  profiles,  with  an
uncertainty of just few eV.\\
The \ion{Si}{14}  and \ion{S}{16} lines present a  small redshift with
respect to  the expected rest-frame  values, while the  other emission
lines present values consistent  with rest-frame measurements. We note
that  the detected  forbidden  lines  are hard  to  be interpreted  as
redshifted resonance  lines, given that the  energy separation between
the forbidden  and the  resonance is  in the range  25--40 eV  for the
He-like ions under consideration,  while the redshift as measured from
the H-like  emission line  is a few  eV.  We cannot,  however exclude,
that  the intercombination  lines  can be  partially  blended with  the
forbidden ones.\\
The widths of the lines (except for the \ion{Mg}{12} Ly$_{\alpha}$ and
Ly$_{\beta}$ that were unconstrained by  the fit, so that we choose to
fix the values  of these parameters to a reference value  of 2 eV) are
generally  broadened  by  a  few  electronvolts. We  show  data,  with
superimposed best-fit model, in  this soft X-ray region (1.4--2.7 keV)
in  the left  panel of  Figure~\ref{fig6}.  The  \ion{Ca}{19} resonant
emission  line,  present during  the  quiescent  state,  is no  longer
detected in the  total average spectrum, probably due  to the stronger
continuum  flux at  this energy,  or, more  probably, by  the combined
effect of emission (during the  hard state) and absorption (during the
flaring  state) seen  in the  time-selected spectra,  but  we consider
highly   significant  the   detection   of  the   forbidden  line   of
\ion{Ca}{19}.  We also  observe in the spectrum an  absorption line at
5.689  $\pm$  0.016 keV,  $\sigma  =  0.013$  (upper value  0.03)  and
equivalent width  of 7 eV, that  is hard to identify  with a resonance
line of any element.  Its  significance is at $\sim$ 3 $\sigma$.  This
feature was not  detected in the quiescent/hard spectra,  while in the
flare integrated  spectrum this feature  is required at  2.8 $\sigma$.
In the 6.4--7.1 keV energy band the spectrum closely follows the shape
of  the spectrum  during the  flaring  episodes, showing  the iron  Fe
K$\alpha$ emission line, a broad absorption feature in the 6.5-6.6 keV
range and the \ion{Fe}{25} and \ion{Fe}{26} resonant absorption lines.
We  also check  for two  expected  further absorption  edges at  fixed
energy position,  corresponding to the expected  edges of \ion{Fe}{25}
(8.828  keV)   and  \ion{Fe}{26}   (9.278  keV),  obtaining   for  the
corresponding optical depth the best-fit  values of 0.2 and 0.6 (upper
values of 0.6  and 1.3 for the He-like  and H-like ion, respectively).
We present at the end  of Table~\ref{tab2} the identified emission and
absorption features  in this case.\\ After having  verified that every
part of the spectrum was  free from any other significant residual, we
obtained a final $\chi^2$ value of 1004 for 1384 degrees of freedom.
%%%%%%%%%%%%%%%%%%%%%%%%%%%%%%%%%%%%%%%%%%%%%%%%%
%%%%%%%%%%%%%%%%%%%%%%%%%%%%%%%%%%%%%%%%%%%%%%%%%
\section{Discussion}

In    the   \textit{Chandra}    observations    performed   in    2000
\citep{schulz02}, the  Cir X-1 spectrum clearly showed  a rich variety
of  emission/absorption  features, the  most  striking  of which  were
resolved P-Cygni profiles of He- and H-like elements. This observation
is  taken  at  a higher  orbital  phase  with  respect to  the  almost
phase-zero observations I and II in Paper I.  Compared to the previous
observations, we  observe in this case two  different spectral states,
within  the same  observational window.   A  low-countrate, quiescent,
state, followed  by more complex, highly variable,  flaring states. We
do not  detect any  P Cygni  profile in the  spectrum, but  we observe
resonant  emission  lines  from H-like  ions  of  Mg,  Si and  S,  the
forbidden lines of He-like Si, S and Ca, the Fe K$\alpha$ fluorescence
line at  $\sim$ 6.4 keV and  a complex variable  absorption pattern in
the Fe K$\alpha$ region during the flaring states of the source.

\subsection{The continuum emission}

During the  quiescent state  the source should  be accreting at  a few
percent of its  Eddington limit (for an assumed distance  of 6 kpc and
for  a  1.4 M$_{\odot}$  NS),  that  is  considerably less  than  what
reported  for  Observation  I  ($\sim$  at the  Eddington  limit)  and
Observation  II ($\sim$  10\%  L$_{Edd}$) in  Paper  I.  Although  the
limited   energy  coverage   and  the   low  counting   statistics  of
\textit{Chandra} can give only a first order approximation to the real
broadband spectrum, we note that the continuum spectrum is not similar
to any  previously observed spectrum  of this source. Its  rather hard
shape is  much closer  to that of  a typical accretion  powered pulsar
\citep{zurita06}; also  the derived blackbody temperature  is too high
to  be taken  as  a  thermalized blackbody  temperature  of the  inner
accreting regions. The primary X-ray source results partially absorbed
by a cold neutral medium, that however, does not completely screen out
the primary X-ray  flux.  The derived column density  of the absorbing
material is not compatible with the (2.3 $\pm$ 0.2) $\times$ 10$^{22}$
cm$^{-2}$ of the  zero-phase observations; we note about  one order of
magnitude  increase during  our  observation, with  also an  increased
value of the covering fraction.  The spectrum in this state also shows
a strong  Fe K$\alpha$  edge, and, although  the value of  the derived
optical  depth cannot  be  well constrained  by  the fit,  as this  is
entangled  with the  derived column  density of  the  partial covering
component,  its increase, with  respect to  the phase  0 observations,
seems  to correlate with  the increase  of the  column density  of the
partial  covering medium,  thus leading  to the  conjecture  that iron
present in the neutral medium is strongly overabundant with respect to
the interstellar abundance.

The lack of  P Cygni profiles in the  spectrum during this observation
could  be related to  the change  in luminosity  of the  primary X-ray
flux:  the   reduced  irradiating  flux  with  respect   to  the  past
\textit{Chandra}  observations  might  not  be  able  to  consistently
evaporate part of the accretion  disk, or, alternatively, the disk may
not be present and matter may  accrete onto the compact object along a
direction that we  do not directly observe (the  absence of any softer
component in the spectrum could  also indicate that the accretion disk
does  not  contribute to  the  X-ray luminosity  and  that  it may  be
truncated at a large distance from the compact object).

The spectral continuum shape has  a dramatic change during the flaring
state.  Previous models considered that the dipping activity in Cir
X-1 could be related to the almost edge-on geometry of the system that
makes  our line-of-sight  directly plunge  in the  turbulent accreting
external layers  of an accretion disk  \citep{shirey99}. This scenario
explains  the  lightcurve variability  as  a  combined  effect of  the
variable accretion  rate and  of the change  in density of  the colder
accreting material.   In this observation the major  difference in the
pre-flaring  and the  flaring states  is the  fraction of  the partial
covering  component  of  the  primary  X-ray flux,  together  with  an
increased  continuum  luminosity.   If  this  is a  consequence  of  a
turbulent higher  mass accretion rate, it  can be argued  that this is
also  the cause  of a  geometrical inflation  of the  occulting medium
giving the increase in the covering fraction.

During the flaring states, the  soft emission lines are more difficult
to detect, given the  high photoelectric absorption at these energies.
However, the total 'flare'  spectrum show that these features continue
to be  present in the spectrum,  growing in intensity  with respect to
the  quiescent/hard   state,  in  correlation  with   a  more  intense
irradiating  continuum  flux.   

\citet{boirin05} analyzing the spectrum of the source 4U 1323-63, in a
dipping and in a persistent  state, found that spectral changing could
be modelled  by variations in  the properties of  a neutral and  of an
ionized  absorber.  We  suggest that  Cir X-1  could also  represent a
paradigmatic example of the  complex interplay of these two absorbers.
Moreover,  in Cir  X-1 there  could also  be a  rapid changing  in the
geometry of  the emitting media; a proof  of that may be  given by the
emission  line of \ion{Fe}{25}  that we  observe during  the quiescent
state and that appears in absorption during the flaring state.

\subsection{Line diagnostics: indications of a hot, photo-ionized, plasma}

The features  that we observe  in emission during the  quiescent state
are from H-like Si, and from He-like  Ca and Fe; we argue that the low
energy  detected emission lines  in the  total integrated  spectrum of
\ion{Mg}{12} (where  we detect the  Lyman $\alpha$, $\beta$  and 
$\gamma$ transitions), \ion{S}{16} and  the forbidden He-like lines of
\ion{Si}{13}  (blended,  however  with the  \ion{Mg}{12}  Ly$\gamma$),
\ion{S}{15}  and \ion{Ca}{19}  could also  be common  features  in the
quiescent/hard  spectrum,  not  revealed   only  because  of  the  low
statistics.  Under this assumption, we  infer that the weakness of the
He-like emission  lines translate  into a rather  higher value  of the
ionization parameter  with respect to  the value reported in  Paper I.
The ionization parameter $\xi=L/n_e  r^2$ (where $n_e$ is the electron
density,  $r$  is  the  distance  from  the  source  and  $L$  is  the
X-luminosity) that  characterizes the  Mg, Si, and  S H-like  ions and
avoids  having  appreciable  fractions of  the corresponding  He-like
ions,  is  above  log($\xi$) $>$  3, in the limit of a low density gas
and a hard ionizing flux, F$_{ion} \propto E^{-1}$, \citealt{kallman01}).

We  consider the  detection of  the forbidden  lines  of \ion{Si}{13},
\ion{S}{15}  and   \ion{Ca}{19}  as  a  strong   confirmation  of  the
photoionized nature of the emitting  plasma.  However we note that the
relative  weakness/absence  of  the  resonance  lines  could  also  be
ascribed to the fact that line resonance photons can be trapped inside
the emitting  plasma, being multiply  emitted and reabsorbed.   If the
thermal electron  temperature is  in the $10^6  - 10^7$ K  range, then
Compton  scattering of the  line photons  will lead  to a  strong line
broadening, thus making them  hard to detect.  Forbidden lines photons
will,  on  the  contrary,  easily  escape the  emitting  medium,  thus
preserving line narrowness.

These lines  appear, at least  the ones that are  statistically better
constrained,  broadened,  the  broadening  being  stronger  for  lower
Z-elements.   Their relative  widths  ($\Delta E  /  E \sim$  (1.5--3)
$\times$ 10$^{-3}$) can be due to a combination of thermal, rotational
and  turbulent  effects.  We  also  note,  a  slight redshift  of  the
brightest lines; we also verified  the common redshift of the emission
lines, and  their common origin, by  leaving as a  free parameter only
the \ion{Si}{14} centroid energy and fixing the other line energies at
the expected rest frame difference (i.e  $E_{ S~XVI } = E_{ Si~XIV } +
0.6162$ keV,  and similarly for  the other H-like emitting  lines), in
the  total  time  integrated  spectrum.   We found  that  the  fit  is
equivalent to  the one, where  the line energies are  free parameters,
all the  emitted lines, if  Doppler redshifted, share a  common radial
velocity of  about $\sim$ 140  km s$^{-1}$.  The significance  of this
result   is   at  at   more   than   2   $\sigma$  confidence   level.
\citet{schulz02}  derived from  the He-like  resonance,  forbidden and
intercombination line of the He-like \ion{Mg}{12} a value G=$F_i+F_z /
F_r = 2.9 \pm 1.0$, where  the $F_i$, $F_z$ and $F_r$ are respectively
the flux values of  the intercombination, forbidden and resonance line
which  strongly  indicates  the  presence  of  a  photoionized  plasma
\citep{purquet00}  as the  proper line  emitting region.   During this
\textit{Chandra} observation, we do  not observe the He-like resonance
lines, while the intercombination lines cannot be clearly resolved, so
that we cannot appropriately use the  R and G diagnostics to infer the
physical properties of the emitting plasma. However, as pointed out by
\citet{purquet00}, the presense of forbidden lines and the weakness of
the resonant ones is by  itself a good indication of the photo-ionized
nature  of the emitting  plasma.\\ Using  the calculations  present in
\citet{purquet00} we  derive from the intensity of  the forbidden line
of  \ion{Si}{13} we derive  an upper  limit to  the plasma  density of
$n_e$  of 10$^{13}$  cm$^{-3}$.  The  intensity of  the  H-like lines,
compared  to  the the  corresponding  He-like  lines,  gives a  second
important clue,  namely that the plasma temperature  is extremely high
(in the range 10$^6$--10$^7$ K) in  order to allow a high ratio of the
H-like ion  population to the  He-like one. The plasma  is, therefore,
dominated by recombination, rather than by collisional excitation.

\subsection{Absence of radiative recombination continua}

Another  characteristic  signature of  a  purely photo-ionized  plasma
would be the detection of the radiative recombination continuum (RCC),
which should appear  in the spectrum as an  emitting line feature just
above  the K  H-like/He-like  ion edge.   We  do not  observe any  RCC
feature   above  the   K  edge   of  \ion{Mg}{12},   \ion{Si}{14}  and
\ion{S}{16},  both in  the quiescent/hard  region and  in  the flaring
state.  If the emitting plasma  is strongly photoionized, we note that
the lack of such features could generally be ascribed to one, or more,
of    these   physical   processes    \citep{bautista98}:   three-body
recombination,   collisional   excitation   or  resonant   fluorescent
excitation.  The first mechanism will operate in a regime of very high
density and low  temperature; however, for very high  Z elements, such
as  Si  and  S,  this   process  is  not  competitive  with  radiative
recombination  (as three-body  recombination  coefficient scales  very
rapidly with nuclear  charge Z, as $Z^{-6}$), and its  impact is to be
considered marginal.   Collisional excitation,  on the contrary,  is a
viable  candidate  process for  populating  $n=2$  levels, opposed  to
radiative recombination,  provided that temperatures are  in excess of
$10^6$ K  and there is  a significant fraction  of H-like ions  in the
ground state, but the presence of forbidden lines indicate that plasma
density is  extremely low to allow collisions  to efficiently operate.
Finally, photo-excitation  is a very  efficient way to  populate upper
levels, but  it imposes  a particular geometry  of the  X-ray emitting
continuum  region and line  emitting region,  i.e.  that  the emitting
photoexcited  plasma is  located away  from  the source  of the  X-ray
continuum  and  is  not  located  between the  X-ray  source  and  the
observer.   Another,  perhaps  simpler,  explanation is  that  RCC  is
undetected because the  plasma is hot enough to  smear out RCC photons
into  the  continuum  (the  broadening  of  this  feature  depends  on
temperature   and   is   stronger   for   higher   temperatures,   see
\citet{liedahl96}).

\subsection{The iron and calcium features}

The presence of  the fluorescence line of the  Fe K$\alpha$ during the
quiescent emission could be due, as in the case of the H-like emission
lines at  lower energies,  to reflection in  the colder border  of the
accretion disk of the primary  X-ray incident flux.  The line could be
broadened by the blending of all the low ionization states of iron (Fe
I -  Fe IX), and  we also observe  that the corresponding edge  has an
edge  energy that  is  slightly higher  than  the rest-frame  expected
threshold  at 7.11  keV, thus  supporting the  idea that  a consistent
fraction  of  the  reflecting   iron  is  in  a  low-ionization  state
\citep{kallman01}.   This  state   shows  only  one  clearly  detected
absorption feature  at $\sim$ 6.49  keV. We take as  highly improbable
the  origin of  this feature  as a  blueshifted Fe  K$\alpha$  line of
nearly neutral photoionized iron, given the required extreme high bulk
velocity of the  absorbing medium ($3700_{-800}^{+1800}$ km s$^{-1}$).
On the  contrary, we are led  to interpret this  feature, according to
the  calculations  of  \citet{kallman04},  as blending  of  absorption
resonant transitions  of \ion{Fe}{20}-\ion{Fe}{24}. This  is the first
observation where resonant blended absorption features of intermediate
ionization states of iron are observed in the spectrum of Cir X--1.

As  in  the case  of  4U  1323-62  \citep{boirin05}, of  MXB  1659-298
\citep{sidoli01},   X    1624-49   \citep{parmar02}   and    GX   13+1
\citep{sidoli02},  we also  find  a broad  line  centered at  energies
between  6.5 and 6.6  keV; however,  the Cir  X-1 spectrum  shows this
feature in absorption,  while in the other sources  it was observed in
emission.   We also  note that  the large  width associated  with this
feature could  also indicate a  rather large gradient in  the velocity
field  of  the absorbing  medium;  the  \ion{Fe}{25} and  \ion{Fe}{26}
lines, which are both better resolved in the total integrated spectrum
are  strongly  broadened  and  are  blueshifted with  respect  to  the
rest-frame  energies.    Together  with  the   absorption  feature  of
\ion{Ca}{20}, these lines  give an average blueshift of  370 $\pm$ 100
km s$^{-1}$ that we interpret as due to a bulk motion of the absorbing
medium.   We also  note that  this value  is significantly  lower with
respect to  the reported past  \textit{Chandra} observations in Paper  I, where
the H-like  and He-like  iron lines showed  a $\sim$ 1000  km s$^{-1}$
blueshift.

If we neglect the filling effect of the emission lines observed during
the quiescent state  and if we make the assumption  that the lines are
not  saturated, we  can derive  a rough  estimate (to  be  more rightly
intended as lower limits) of the column densities of the absorbing hot
cloud from the absorption lines  of H-like and He-like iron, using the
calcultions of \citet{spitzer78}, with the relation:
\begin{displaymath}
\frac{W_{\lambda}}{\lambda} = \frac{\pi e^2}{m_e c^2} N_j \lambda f_{ij}=
8.85 \times 10^{-13} N_j \lambda f_{ij}
\end{displaymath} 
where N$_j$ is  the column density for the  relevant species, $f_{ij}$
is the  oscillator strenght and $W_{\lambda}$ the  equivalent width of
the line  and $\lambda$ is  the wavelenght expressed in  cm.

Adopting  f$_{ij}$=0.798  and   f$_{ij}$=0.416  for  \ion{Fe}{25}  and
\ion{Fe}{26}   respectively  \citep{verner95}  and   the  best-fitting
paramenters  of  Table~\ref{tab2}, we  derived  from the  best-fitting
parameters a column density  of $\sim$ 2.3$\times$ 10$^{17}$ cm$^{-2}$
and 8.3$\times$ 10$^{17}$  cm$^{-2}$ for \ion{Fe}{25} and \ion{Fe}{26}
respectively. This rough  estimation allows us to derive  the ratio of
the ionization fractions  of the two ion species  ($\sim$ 3.6), and to
assign  a  log$\xi$ value  for  the  ionization  parameter $\sim$  4.2
\citep{kallman01}.\\  The ratio  of the  H-like/He-like  abundances of
calcium suffer  for larger uncertainties,  but still the value  of the
ionization  parameter that we  derived (log$\xi$  $\sim$ 3.5  from the
best-fitting   values),  confirms   the  presence   of   an  extremely
photoionized  plasma.  However, we  note that  to obtain  the observed
equivalent  width  ratio  Ca   XX/Fe  XXVI  (neglecting  any  possible
saturation  effect  on the  lines),  we  infer  that calcium  must  be
overabundant  by almost  an order  of  magnitude with  respect to  the
cosmic abundances.

The absorption edge that is detected  above 7.0 keV, at an energy that
is considerably  lower than  the 7.11 keV  expected value, for  the Fe
K$\alpha$  edge, could  result from  the radiative  and  Auger damping
effects after the creation of the K-hole \citep[see also][]{palmeri02,
  ross96}.   As  predicted  by  the  high  ionization  parameter,  the
resonances are  spread over more than  500 eV width,  although we have
also to  take into  account that the  strong neutral Fe  edge observed
during the  hard state certainly introduces a  significant fraction of
fictitious "smearing" in the detected edge.

\subsection{A highly redshifted absorption line ?}
The detection  at 3.1 $\sigma$  significance of an absorption  line at
5.689 keV  could be due to  an extremely redshifted  (z $\simeq$ 0.18)
\ion{Fe}{26} line, produced in  the photosphere of the compact object.
We derive, for a 1.4 M$_{\odot}$ NS a distance from the compact object
of $\sim$  14 km,  if the line  is purely  gravitationally redshifted,
excluding any possible Doppler effect. The absence of substructures in
this  line, if  the  line  is rotationally  broadened,  would imply  a
rotational frequency  $\geq$ 50 Hz  \citep{chang06}. We note  that the
presence of this feature, with a slight shift in energy, has also been
reported in \citet{iaria07}, and a  similar feature has been also been
observed by \citet{longinotti07}  in the spectrum of the  AGN Mrk 335.
In the latter  case the redshift was interpreted as  due to an optical
thick radial inflow of the accreting matter near the compact object at
$\sim$ 0.11 -  0.15 c velocity.  Such explanation  could also apply to
the case  of Cir  X-1, and  we computed in  this case,  neglecting the
effects of the radiation pressure  on the infalling matter, a distance
to  the   central  object  of  $\sim$  60   R$_g$.   However,  further
observations are required to confirm this detection.

\subsection{System geometry}

In the following we explore and discuss two possible geometrical scenarios
for the system.

The  presence of  highly  ionized absorption  features during  flaring
episodes, together  with emission  features from H-like  ions requires
two spatially distinct zones; one optically thick region, with a large
$\xi$   gradient,   rapidly    varying   according   to   the   global
luminosity/accretion  rate of  the source,  owing also  large gradient
velocity fields;  and a  second zone, optically  thinner, photoionized
plasma that  is present  during both the  persistent emission  and the
flaring episodes.  \citet{jimenez02} calculated  the effects of a hard
X-ray source illuminating an accretion disk; it is shown that above an
assumed Shakura-Sunyaev  disk \citep{shakura73} and  under hydrostatic
balance,  there is  a natural  split into  a two-zone  medium:  a warm
corona,   or   atmosphere,   where   photoionization   and   radiative
recombination  balance energetically  each  other, and  a hot  corona,
dominated by  the balance of  Compton heating and cooling.   The total
volume, ionization fraction and density in the two zone determines the
strength of  the emission/absorption  features impressed in  the total
spectrum.   The weakness  of  He-like ions  lines  during our  \textit{Chandra}
observation  (with  the only  exception  of  the He-like  \ion{Ca}{19}
emission  line  in  the  hard  state), compared  to  previous  \textit{Chandra}
observations,  could be  traced back  in  the interplay  of these  two
zones.  If  the He-like emitting  region is the colder  atmosphere and
the  H-like emitting region  is set  inside the  hot corona,  than the
ratio of the H-like/He like emission lines indicates a different ratio
of  the emitting  volumes, given  also that  the  ionization structure
above the  photosphere does not  sensibly depend on the  primary X-ray
luminosity.   If   a  thermal  instability  is   operating,  then  our
observation can  be in  agreement with a  condensing solution  for the
disk atmosphere.  Such scenario has, however, two open problems: where
to  locate the  optically thicker  absorbing medium  during  the flare
episodes and how to account for the redshift of the emitting lines and
the  blueshift  of  the  absorption  ones.  As  also  pointed  out  by
\citet{jimenez02},  the  presence  of  strong outflows/inflows  in  an
accreting  system  breaks  up  the  assumption  of  local  hydrostatic
balance, and  the solutions for the  disk atmosphere can  offer only a
partial, or  maybe incorrect, modelling.   We can only argue  that the
absorbing  plasma,  for the  required  high  ionization parameter  and
density, can be located in the inner part of the accretion flow, where
a strong magnetic field prevents the flow from entering the inner part
of the system where it would disrupt the accretion disk.  Part of this
hot,  optically  thick matter  can  eventually  be  expelled from  the
system, during periods of turbulent enhanced accretion, at this radius
causing the observed blueshift.  On the contrary, if the emitted lines
are  produced inside a  disk corona  the redshift  is caused  again by
Doppler motion of the emitting plasma, dragged towards the accretor.

Alternatively,  we  point  out  a  second viable  scenario,  that  has
recently emerged \citep{iaria07}. Emission  features occur in a beamed
outflow, that could be seen as  the X-ray counterpart of the radio jet
observed  by \citet{fender04}, the  intensity of  the redshift  of the
lines depends  on the inclination of  the outflow with  respect to our
line of sight; absorption features are located, on the contrary, in an
optically  thick wind-driven  corona; the  spectrum is  signed  by the
sampling  of the  primary continuum  into regions  characterized  by a
gradient  of  the  ionization  parameter, optical  depth  and  outflow
velocity  field.   The passage  from  the  quiescient  to the  flaring
activity causes  an a geometrical inflation  of the wind,  that at the
outer  radius completely  screens the  central X-ray  emitting object,
while  in the  inner regions  gives  the resonant  features of  highly
ionized  elements. Given the geometrical constraint of the get geometry
\citep{fender04,tudose06}, this  scenario, however,  requires that  the system
geometry  is not  perfectly  edge-on and  contradicts the  theoretical
expectation of an optical thin plasma above the disk.
\subsection{Conclusions}
We analyzed \textit{Chandra} HETGS data of Cir X-1 during phase passage 0.223-0.261.
Based on the flux variability, the source clearly showed a complex
spectral changing both in the continuum and in the features. This observation
reveals the following, partly new, aspects on this source:
\begin{itemize}
\item The passage from a  quiescent to a flaring activity determines a
  change both in  the luminosity and in the  covering fraction of
  the  absorbing  medium,  that  during the  flare  almost  completely
  screens out the source of the continuum emission.
\item During flare episodes the 6.4--7.0 keV region is dominated by broad
absorption lines of moderately and highly ionized iron ions, that cover a large
range of possible ionization factors.
\item During flaring the source still displays emission lines, whose intensity
increases with increased continuum luminosity.
\item Emission lines stem from a purely photo-ionized medium with  temperatures
 in excess of 10$^6$ K and electron density below $n_e \sim$  10$^{13}$ cm$^{-3}$. 
\item We found a  possible detection of a strong redshifted 
\ion{Fe}{26} absorption line at z=0.18.    
\end{itemize}
\acknowledgements This work was partially supported by the
Italian Space  Agency (ASI) and the Ministero dell'
Universit\'a e della Ricerca (MiUR).
%\bibliographystyle{apj}
%\bibliography{references}

\begin{thebibliography}{47}
\expandafter\ifx\csname natexlab\endcsname\relax\def\natexlab#1{#1}\fi

\bibitem[{{Bautista} {et~al.}(1998){Bautista}, {Kallman}, {Angelini},
  {Liedahl}, \& {Smits}}]{bautista98}
{Bautista}, M.~A., {Kallman}, T.~R., {Angelini}, L., {Liedahl}, D.~A., \&
  {Smits}, D.~P. 1998, \apj, 509, 848

\bibitem[{{Belloni} {et~al.}(2007){Belloni}, {M{\'e}ndez}, \&
  {Homan}}]{belloni07}
{Belloni}, T., {M{\'e}ndez}, M., \& {Homan}, J. 2007, \mnras, 376, 1133

\bibitem[{{Boirin} {et~al.}(2005){Boirin}, {M{\'e}ndez}, {D{\'{\i}}az Trigo},
  {Parmar}, \& {Kaastra}}]{boirin05}
{Boirin}, L., {M{\'e}ndez}, M., {D{\'{\i}}az Trigo}, M., {Parmar}, A.~N., \&
  {Kaastra}, J.~S. 2005, \aap, 436, 195

\bibitem[{{Boutloukos} {et~al.}(2006){Boutloukos}, {van der Klis},
  {Altamirano}, {Klein-Wolt}, {Wijnands}, {Jonker}, \& {Fender}}]{boutloukos06}
{Boutloukos}, S., {van der Klis}, M., {Altamirano}, D., {Klein-Wolt}, M.,
  {Wijnands}, R., {Jonker}, P.~G., \& {Fender}, R.~P. 2006, \apj, 653, 1435

\bibitem[{{Brandt} {et~al.}(1996){Brandt}, {Fabian}, {Dotani}, {Nagase},
  {Inoue}, {Kotani}, \& {Segawa}}]{brandt96}
{Brandt}, W.~N., {Fabian}, A.~C., {Dotani}, T., {Nagase}, F., {Inoue}, H.,
  {Kotani}, T., \& {Segawa}, Y. 1996, \mnras, 283, 1071

\bibitem[{{Chang} {et~al.}(2006){Chang}, {Morsink}, {Bildsten}, \&
  {Wasserman}}]{chang06}
{Chang}, P., {Morsink}, S., {Bildsten}, L., \& {Wasserman}, I. 2006, \apjl,
  636, L117

\bibitem[{{Clark} {et~al.}(1975){Clark}, {Parkinson}, \& {Caswell}}]{clark75}
{Clark}, D.~H., {Parkinson}, J.~H., \& {Caswell}, J.~L. 1975, \nat, 254, 674

\bibitem[{{Clark} {et~al.}(2003){Clark}, {Charles}, {Clarkson}, \&
  {Coe}}]{clark03}
{Clark}, J.~S., {Charles}, P.~A., {Clarkson}, W.~I., \& {Coe}, M.~J. 2003,
  \aap, 400, 655

\bibitem[{{Clarkson} {et~al.}(2004){Clarkson}, {Charles}, \&
  {Onyett}}]{clarkson04}
{Clarkson}, W.~I., {Charles}, P.~A., \& {Onyett}, N. 2004, \mnras, 348, 458

\bibitem[{{Ebisawa}(1991)}]{ebisawa91}
{Ebisawa}, K. 1991, PhD thesis, , Univ.~Tokyo; ISAS Research Note No.~483,
  (1991)

\bibitem[{{Fender} {et~al.}(1998){Fender}, {Spencer}, {Tzioumis}, {Wu}, {van
  der Klis}, {van Paradijs}, \& {Johnston}}]{fender98}
{Fender}, R., {Spencer}, R., {Tzioumis}, T., {Wu}, K., {van der Klis}, M., {van
  Paradijs}, J., \& {Johnston}, H. 1998, \apjl, 506, L121

\bibitem[{{Fender} {et~al.}(2004){Fender}, {Wu}, {Johnston}, {Tzioumis},
  {Jonker}, {Spencer}, \& {van der Klis}}]{fender04}
{Fender}, R., {Wu}, K., {Johnston}, H., {Tzioumis}, T., {Jonker}, P.,
  {Spencer}, R., \& {van der Klis}, M. 2004, \nat, 427, 222

\bibitem[{{Glass}(1994)}]{glass94}
{Glass}, I.~S. 1994, \mnras, 268, 742

\bibitem[{{Haynes} {et~al.}(1978){Haynes}, {Jauncey}, {Murdin}, {Goss},
  {Longmore}, {Simons}, {Milne}, \& {Skellern}}]{haynes78}
{Haynes}, R.~F., {Jauncey}, D.~L., {Murdin}, P.~G., {Goss}, W.~M., {Longmore},
  A.~J., {Simons}, L.~W.~J., {Milne}, D.~K., \& {Skellern}, D.~J. 1978, \mnras,
  185, 661

\bibitem[{{Haynes} {et~al.}(1986){Haynes}, {Komesaroff}, {Jauncey}, {Caswell},
  {Milne}, {Kesteven}, {Wellington}, \& {Preston}}]{haynes86}
{Haynes}, R.~F., {Komesaroff}, M.~M., {Jauncey}, D.~L., {Caswell}, J.~L.,
  {Milne}, D.~K., {Kesteven}, M.~J., {Wellington}, K.~J., \& {Preston}, R.~A.
  1986, \nat, 324, 233

\bibitem[{{Heinz} {et~al.}(2007){Heinz}, {Schulz}, {Brandt}, \&
  {Galloway}}]{heinz07}
{Heinz}, S., {Schulz}, N.~S., {Brandt}, W.~N., \& {Galloway}, D.~K. 2007,
  \apjl, 663, L93

\bibitem[{{Iaria} {et~al.}(2001{\natexlab{a}}){Iaria}, {Burderi}, {Di Salvo},
  {La Barbera}, \& {Robba}}]{iaria01b}
{Iaria}, R., {Burderi}, L., {Di Salvo}, T., {La Barbera}, A., \& {Robba}, N.~R.
  2001{\natexlab{a}}, \apj, 547, 412

\bibitem[{{Iaria} {et~al.}(2001{\natexlab{b}}){Iaria}, {Di Salvo}, {Burderi},
  \& {Robba}}]{iaria01a}
{Iaria}, R., {Di Salvo}, T., {Burderi}, L., \& {Robba}, N.~R.
  2001{\natexlab{b}}, \apj, 561, 321

\bibitem[{{Iaria}(2007)}]{iaria07}
{Iaria}, R. e.~a. 2007, ApJ submitted

\bibitem[{{Jimenez-Garate} {et~al.}(2002){Jimenez-Garate}, {Raymond}, \&
  {Liedahl}}]{jimenez02}
{Jimenez-Garate}, M.~A., {Raymond}, J.~C., \& {Liedahl}, D.~A. 2002, \apj, 581,
  1297

\bibitem[{{Jonker} {et~al.}(2007){Jonker}, {Nelemans}, \& {Bassa}}]{jonker07}
{Jonker}, P.~G., {Nelemans}, G., \& {Bassa}, C.~G. 2007, \mnras, 374, 999

\bibitem[{{Kallman} \& {Bautista}(2001)}]{kallman01}
{Kallman}, T., \& {Bautista}, M. 2001, \apjs, 133, 221

\bibitem[{{Kallman} {et~al.}(2004){Kallman}, {Palmeri}, {Bautista}, {Mendoza},
  \& {Krolik}}]{kallman04}
{Kallman}, T.~R., {Palmeri}, P., {Bautista}, M.~A., {Mendoza}, C., \& {Krolik},
  J.~H. 2004, \apjs, 155, 675

\bibitem[{{Kaluzienski} {et~al.}(1976){Kaluzienski}, {Holt}, {Boldt}, \&
  {Serlemitsos}}]{kaluzienski76}
{Kaluzienski}, L.~J., {Holt}, S.~S., {Boldt}, E.~A., \& {Serlemitsos}, P.~J.
  1976, \apjl, 208, L71

\bibitem[{{Liedahl} \& {Paerels}(1996)}]{liedahl96}
{Liedahl}, D.~A., \& {Paerels}, F. 1996, \apjl, 468, L33+

\bibitem[{{Longinotti} {et~al.}(2007){Longinotti}, {Sim}, {Nandra}, \&
  {Cappi}}]{longinotti07}
{Longinotti}, A.~L., {Sim}, S.~A., {Nandra}, K., \& {Cappi}, M. 2007, \mnras,
  374, 237

\bibitem[{{Mignani} {et~al.}(2002){Mignani}, {De Luca}, {Caraveo}, \&
  {Mirabel}}]{mignani02}
{Mignani}, R.~P., {De Luca}, A., {Caraveo}, P.~A., \& {Mirabel}, I.~F. 2002,
  \aap, 386, 487

\bibitem[{{Morrison} \& {McCammon}(1983)}]{morrison83}
{Morrison}, R., \& {McCammon}, D. 1983, \apj, 270, 119

\bibitem[{{Palmeri} {et~al.}(2002){Palmeri}, {Mendoza}, {Kallman}, \&
  {Bautista}}]{palmeri02}
{Palmeri}, P., {Mendoza}, C., {Kallman}, T.~R., \& {Bautista}, M.~A. 2002,
  \apjl, 577, L119

\bibitem[{{Parkinson} {et~al.}(2003){Parkinson}, {Tournear}, {Bloom}, {Focke},
  {Reilly}, {Wood}, {Ray}, {Wolff}, \& {Scargle}}]{saz03}
{Parkinson}, P.~M.~S., {Tournear}, D.~M., {Bloom}, E.~D., {Focke}, W.~B.,
  {Reilly}, K.~T., {Wood}, K.~S., {Ray}, P.~S., {Wolff}, M.~T., \& {Scargle},
  J.~D. 2003, \apj, 595, 333

\bibitem[{{Parmar} {et~al.}(2002){Parmar}, {Oosterbroek}, {Boirin}, \&
  {Lumb}}]{parmar02}
{Parmar}, A.~N., {Oosterbroek}, T., {Boirin}, L., \& {Lumb}, D. 2002, \aap,
  386, 910

\bibitem[{{Porquet} \& {Dubau}(2000)}]{purquet00}
{Porquet}, D., \& {Dubau}, J. 2000, \aaps, 143, 495

\bibitem[{{Ross} {et~al.}(1996){Ross}, {Fabian}, \& {Brandt}}]{ross96}
{Ross}, R.~R., {Fabian}, A.~C., \& {Brandt}, W.~N. 1996, \mnras, 278, 1082

\bibitem[{{Schulz} \& {Brandt}(2002)}]{schulz02}
{Schulz}, N.~S., \& {Brandt}, W.~N. 2002, \apj, 572, 971

\bibitem[{{Shakura} \& {Syunyaev}(1973)}]{shakura73}
{Shakura}, N.~I., \& {Syunyaev}, R.~A. 1973, \aap, 24, 337

\bibitem[{{Shirey} {et~al.}(1999){Shirey}, {Levine}, \& {Bradt}}]{shirey99}
{Shirey}, R.~E., {Levine}, A.~M., \& {Bradt}, H.~V. 1999, \apj, 524, 1048

\bibitem[{{Sidoli} {et~al.}(2001){Sidoli}, {Oosterbroek}, {Parmar}, {Lumb}, \&
  {Erd}}]{sidoli01}
{Sidoli}, L., {Oosterbroek}, T., {Parmar}, A.~N., {Lumb}, D., \& {Erd}, C.
  2001, \aap, 379, 540

\bibitem[{{Sidoli} {et~al.}(2002){Sidoli}, {Parmar}, {Oosterbroek}, \&
  {Lumb}}]{sidoli02}
{Sidoli}, L., {Parmar}, A.~N., {Oosterbroek}, T., \& {Lumb}, D. 2002, \aap,
  385, 940

\bibitem[{{Spitzer}(1978)}]{spitzer78}
{Spitzer}, L. 1978, {Physical processes in the interstellar medium} (New York
  Wiley-Interscience, 1978.~333 p.)

\bibitem[{{Stewart} {et~al.}(1993){Stewart}, {Caswell}, {Haynes}, \&
  {Nelson}}]{stewart93}
{Stewart}, R.~T., {Caswell}, J.~L., {Haynes}, R.~F., \& {Nelson}, G.~J. 1993,
  \mnras, 261, 593

\bibitem[{{Stewart} {et~al.}(1991){Stewart}, {Nelson}, {Penninx}, {Kitamoto},
  {Miyamoto}, \& {Nicolson}}]{stewart91}
{Stewart}, R.~T., {Nelson}, G.~J., {Penninx}, W., {Kitamoto}, S., {Miyamoto},
  S., \& {Nicolson}, G.~D. 1991, \mnras, 253, 212

\bibitem[{{Tauris} {et~al.}(1999){Tauris}, {Fender}, {van den Heuvel},
  {Johnston}, \& {Wu}}]{tauris99}
{Tauris}, T.~M., {Fender}, R.~P., {van den Heuvel}, E.~P.~J., {Johnston},
  H.~M., \& {Wu}, K. 1999, \mnras, 310, 1165

\bibitem[{{Tennant} {et~al.}(1986){Tennant}, {Fabian}, \& {Shafer}}]{tennant86}
{Tennant}, A.~F., {Fabian}, A.~C., \& {Shafer}, R.~A. 1986, \mnras, 221, 27P

\bibitem[{{Tudose} {et~al.}(2006){Tudose}, {Fender}, {Kaiser}, {Tzioumis}, {van
  der Klis}, \& {Spencer}}]{tudose06}
{Tudose}, V., {Fender}, R.~P., {Kaiser}, C.~R., {Tzioumis}, A.~K., {van der
  Klis}, M., \& {Spencer}, R.~E. 2006, \mnras, 372, 417

\bibitem[{{Verner} {et~al.}(1996){Verner}, {Ferland}, {Korista}, \&
  {Yakovlev}}]{verner96}
{Verner}, D.~A., {Ferland}, G.~J., {Korista}, K.~T., \& {Yakovlev}, D.~G. 1996,
  \apj, 465, 487

\bibitem[{{Verner} \& {Yakovlev}(1995)}]{verner95}
{Verner}, D.~A., \& {Yakovlev}, D.~G. 1995, \aaps, 109, 125

\bibitem[{{Zurita Heras} {et~al.}(2006){Zurita Heras}, {de Cesare}, {Walter},
  {Bodaghee}, {B{\'e}langer}, {Courvoisier}, {Shaw}, \& {Stephen}}]{zurita06}
{Zurita Heras}, J.~A., {de Cesare}, G., {Walter}, R., {Bodaghee}, A.,
  {B{\'e}langer}, G., {Courvoisier}, T.~J.-L., {Shaw}, S.~E., \& {Stephen},
  J.~B. 2006, \aap, 448, 261

\end{thebibliography}

\clearpage

%%%%%%%%%%%%%%%%%%%%%%%%%%%%%%%%%%%%%%%%%%%%%%%%%
\begin{deluxetable}{llrrr}
\tabletypesize{\small}
\tablewidth{0pt}
\tablecolumns{6}
\tablecaption{The Continuum Emission}
\tablehead{
\colhead{Parameter} & 
\colhead{Units} & 
\colhead{Hard state} & 
\colhead{Flaring state} & 
\colhead{Integrated} \\ }
\startdata
kT$_{BB}$ & keV       &
$3.07^{+0.7}_{-0.1}$   & 
$1.68 \pm 0.03$ &  
$2.19^{+0.15}_{-0.11}$ \\

L$_{BB}$ & 10$^{36}$ erg / s & 
$2.88 \pm 0.07$  & 
$7.5 \pm 0.2$    & 
$4.68 \pm 0.36$ \\

pcfabs N$_H$ & 10$^{22}$   & 
$16^{+5}_{-3}$      & 
$9.2 \pm 0.3$ & 
$9.0^{+0.5}_{-0.7}$ \\

pcfabs cvr frac &          & 
$0.69^{+0.07}_{-0.06}$ & 
$0.974 \pm 0.005$        & 
$0.93 \pm 0.01$ \\

SmedgeE        & keV  & 
$7.16^{+0.07}_{-0.03}$  & 
$7.13^{+0.13}_{-0.06}$  & 
$7.0^{+0.04}_{-0.11}$ \\

Smedge $\tau$ &      & 
$1.2^{+0.3}_{-0.4}$     & 
$0.9^{+0.1}_{-0.2}$          & 
$1.24^{+1.0}_{-0.25}$ \\
Smedge width &      &   
$0^{+0.2}_{}$                   &                         
$0.42^{+0.4}_{-0.15}$                   &    
$0.54^{+0.8}_{-0.18}$ \\
\hline
$\chi^2/dof$ &      & 
160/316        & 
749/1078       & 
1004/1384   \\
\enddata  
\tablecomments {\footnotesize{Fitting results  for the continuum model
    in the Cir  X--1 HEG/MEG coadded spectra. Data  have been rebinned
    up that a minimum of 25 counts/channel. Uncertainties are reported
    at 90\% confidence  level for a single parameter;  upper limits at
    95\% confidence level.}  }
\label{tab1}
\end{deluxetable}

\begin{deluxetable}{l r r r r r}
\tabletypesize{\small}
\tablewidth{0pt}
\tablecolumns{6}
\tablecaption{Cir X--1 emission and absorption lines}
\tablehead{\colhead{Ion}       &\colhead{Measured E}   &\colhead{ Predicted  E
\tablenotemark{a}} 
               &\colhead{Width}             & 
           \colhead{Eq. Width}           &\colhead{Flux ($\times 10^{-4}$)}\\
           \colhead{}    &\colhead{(keV)}  & \colhead{(keV)}&    
            \colhead{(eV)}& \colhead{(eV)} &  \colhead{(photons cm$^{-2}$ s$^{-1}$)}} 

\startdata
\multicolumn{6}{c}{Hard state} \\
%\hline
\ion{Si}{14}  & $2.0046 \pm 0.0019$ & 2.0054 
              & $3^{+4}_{-3} $ & $57_{-36}^{+43}$ & $1.8^{+0.9}_{-0.8}$\\

\ion{Ca}{19}  &  $3.889^{+0.013}_{-0.009}$  & 3.902 
              & $8^{+17}_{-8}$ & $17_{-11}^{+10}$     &  $ 0.7^{+0.5}_{-0.6} $   \\

\ion{Fe}{1}  & $ 6.403 \pm 0.016    $ & 6.400    
             & $24^{+22}_{-14}$ & $26 \pm 20$  & $1.1 \pm 0.5 $ \\

Fe blended  & $6.49 \pm 0.03$    &  \nodata
              & $50^{+40}_{-30}$  & $-30_{-37}^{+20}$ & $-2.0^{+0.9}_{-1.3}$ \\

\ion{Fe}{25}  & $6.679^{+0.011}_{-0.015}$    & 6.700 
              & $11^{+16}_{-11} $  & $-33 \pm 25$ & $1.4 \pm 0.6$\\
%\hline
\multicolumn{6}{c}{Flaring state (sum)} \\
%\hline
\ion{Si}{14}  & $2.0043$ & 2.0054 
              & $2_{-2}^{+6}$ & $35 \pm 10$ & $14^{+13}_{-8}$\\

\ion{S}{16}  &  $2.6181$ & 2.6216 
             & $4.2$ & $7_{-7}^{+10}$ & $3^{+4}_{-3}$ \\ 

\ion{Ca}{19}  &  $3.911^{+0.005}_{-0.017}$  & 3.902 
              &  $3^{+12}_{-3}$ & $-7_{-50}^{+4}$     &  $-2^{+0.8}_{-17} $   \\

\ion{Ca}{20}  & $4.1065 \pm 0.0025$  &  4.1049 
              & $3^{+4}_{-3} $        & $-10_{-2}^{+3}$    &  $ -3.0 \pm 0.9 $   \\

\ion{Fe}{1}  & $6.398$ & 6.400    
             & $16$ & $15 \pm 13$  & $2.4 \pm 2.0$ \\

Fe blended    & $6.568 \pm 0.020$    &    \nodata
              & $84^{+17}_{-22}$  & $-125_{-25}^{+35}$ & $-23 \pm 5$ \\

\ion{Fe}{25}  & $6.701^{+0.013}_{-0.010}$    & 6.700
              & $11^{+33}_{-11}$   &  $-23_{-15}^{+18}$  & $-3.2^{+1.8}_{-5} $\\

\ion{Fe}{26}  & $6.972^{+0.013}_{-0.009}$  & 6.966 
              & $13^{+18}_{-13}$   & $-36_{-10}^{+13}$  & $-5.5^{+1.9}_{-2.6}$ \\
%\hline
%\cutinhead{{Integrated spectrum}\\
\multicolumn{6}{c}{Integrated spectrum} \\
%\hline
\ion{Mg}{12} (Ly$\alpha$)  & $1.4717^{+0.0025}_{-0.0021}$      & 1.4723     
              & 2   & $40 \pm 20$           & $4.3\pm 2.2$  \\
              
\ion{Mg}{12} (Ly$\beta$) & $1.748 \pm 0.003$ & 1.745 
              & 2   &  $18 \pm 10$  & $2.1^{+8.0}_{-1.9}$ \\

\ion{Mg}{12} (Ly$\gamma$)\tablenotemark{b} &  $1.8395 \pm 0.0015$ & 1.8400  
              & $5.3^{+1.5}_{-2.0}$ & $19^{+5}_{-9}$  & $2.4^{+0.6}_{-1.1}$\\

\ion{Si}{14}  & $2.0043 \pm 0.0011$ & 2.0054 
              & $3.5 \pm 1.1$ & $64^{+7}_{-12}$ & $8.1^{+1.9}_{-1.7}$\\

\ion{S}{15} f & 2.4307 & 2.4307 
              & $2$    & $10^{+6}_{-8}$ & $1.3^{+1.9}_{-1.7}$\\

\ion{S}{16}  &  $2.618 \pm 0.003$ & 2.6216 
             & $4.2 \pm 3.1$ & $22 \pm 9$ & $3.1^{+1.3}_{-1.1}$ \\ 

\ion{Fe}{1}  & $ 6.398 \pm 0.009   $ & 6.400 
             & $16^{+11}_{-16}$ & $26^{+20}_{-12}$  & $2.3^{+1.7}_{-1.0}$ \\

\ion{Ca}{19} f  &   3.8612   & 3.8612
              &   2        & $3.2^{+2.0}_{-1.7}$      & $0.4 \pm 0.3$   \\

\ion{Ca}{20}  &  $4.109 \pm 0.003$  & 4.105 
              & $1.1^{+4.0}_{-1.1}$ & $-6.0 \pm 1.8$      & $-0.82 \pm 0.23$   \\

Fe blended    & $6.541 \pm  0.015$    &  \nodata
              & $95 \pm 20$  & $-125^{+24}_{-20}$  & $-13.1^{+2.4}_{-2.6}$ \\

\ion{Fe}{25}  & $6.713 \pm 0.012$    & 6.7002 
              & $18 \pm 15$  & $-20 \pm 10$  & $-1.6^{+0.8}_{-0.9}$\\

\ion{Fe}{26}  & $ 6.979 \pm 0.008$  & 6.966 
              & $25 \pm 12$   & $-38 \pm 10$  & $-2.3 \pm 1.0$ \\
\enddata  
\tablecomments{Fitting   results   from  emission   and
      absorption  lines   in  the  Cir  X--1  HEG   and  MEG  spectra.
      Uncertainties are reported at 90\% confidence level for a single
      parameter;   upper   limits   at   95\%   confidence   level.}
\tablenotetext{a} {The predicted values  of the detected line energies
  are reported by \citet{verner96}}
\tablenotetext{b}{The line is very
  probably blended with the forbidden line of \ion{Si}{13}}
\label{tab2}
\end{deluxetable}

\begin{deluxetable}{llrrr}
\tabletypesize{\small}
\tablewidth{0pt}
\tablecolumns{5}
\tablecaption{The Flare States}
\tablehead{
\colhead{Parameter} & 
\colhead{Units} & 
\colhead{Flare 1} & 
\colhead{Flare 2} & 
\colhead{Flare 3} \\ }
\startdata
kT$_{BB}$              & keV                      & $1.60^{+0.09}_{-0.05}$    & $1.56^{+0.36}_{-0.11}$           & $1.58^{+0.13}_{-0.05}$           \\
L$_{bb}$               & 10$^{36}$ erg / s        & $6.44 \pm 0.14$          & $8.49 \pm 0.36$               & $9.4 \pm 0.5$                  \\
pcfabs N$_H$ & 10$^{22}$                          & $14.7^{+1}_{-0.7}$        & $10.7^{+1.2}_{-1.6}$           & $7.0^{+1.4}_{-0.6}$             \\
pcfabs cvr frac &                                 & $0.939^{+0.014}_{-0.012}$ & $0.96 \pm 0.03$               & $0.98^{+0.02}_{-0.01}$         \\
Smedge        & keV                               & $7.14^{+0.18}_{-0.21}$    & $7.41^{+0.13}_{-0.3}$          & $7.2^{+0.4}_{-0.1}$            \\
Smedge $\tau$ &                                   & $0.60^{+0.16}_{-0.3}$     & $0.5^{+0.9}_{-0.3}$            & $0.9^{+1.8}_{-0.5}$             \\
Smedge width  & keV                               & $0.12^{+0.6}_{-0.09}$     & $0^{+9}_{}$                   & $0.4^{+5}_{-0.4}$                                \\
%\multicolumn{2}{c}{} & \multicolumn{3}{c}{Absorption Lines} \\
\ion{Ca}{19}  E         & keV              &       3.902                     & \nodata & $3.915^{+0.027}_{-0.023}$    \\
\ion{Ca}{19}  $\sigma$  & eV               &       $22^{+10}_{-14} $          & \nodata &  $30^{+43}_{-11} $   \\
\ion{Ca}{19}  Flux      & $\times 10^{-4}$ &        $ -3.9^{+2.5}_{-2.6} $      & \nodata &  $-7^{+4}_{-6}$  \\
\ion{Ca}{20}  E         & keV               &      $4.107$                   &  $4.107$               & $4.106^{+0.006}_{-0.006}$  \\                
\ion{Ca}{20}  $\sigma$  & eV                &      $0^{+30}_{} $              & $0^{+25}_{} $          &  $0^{+11}_{} $  \\             
\ion{Ca}{20}  Flux      & $\times 10^{-4}$  &       $ -2.3 \pm 1.4 $      & $ -4.5^{+2.4}_{-3.4} $ & $-3.7^{+1.8}_{-1.8}$    \\   
Fe blended  E         & keV  &  $6.58^{+0.02}_{-0.02}$     & $6.59 \pm 0.04$ &   $6.57^{+0.03}_{-0.04}$    \\                 
Fe blended  $\sigma$  & eV  &    $90^{+17}_{-14}$          &  $100^{-30}_{+50}$       & $70 \pm 30$   \\          
Fe blended  Flux    & $\times 10^{-4}$ & $-23 \pm 4$  & $-31^{+9}_{-12}$       & $-19^{+7}_{-6}$  \\   
\ion{Fe}{25} E         & keV            & \nodata  &\nodata  & $6.700$  \\   
\ion{Fe}{25} $\sigma$  & eV             & \nodata  & \nodata & $0^{+23}_{} $  \\   
\ion{Fe}{25} Flux    & $\times 10^{-4}$  &\nodata &\nodata  &  $-5 \pm 3 $  \\   
\ion{Fe}{26} E         & keV  & $6.99^{+0.15}_{-0.04}$       & $6.979^{+0.028}_{-0.022}$   &  $6.972^{+0.08}_{-0.017}$  \\   
\ion{Fe}{26} $\sigma$  & eV  & $32^{+38}_{-20} $             & $8^{-8}_{+70}$             & $6^{+60}_{-6}$  \\   
\ion{Fe}{26} Flux      & $\times 10^{-4}$ & $-5.4 \pm 3 $    & $-9^{+5}_{-18}$            & $-4.5^{+3}_{-6}$   \\ 
\hline
$\chi^2/dof$ &                  & 303/496                & 190/364                     & 279/533                         \\
\enddata  
\tablecomments {\footnotesize{Fitting results  for the continuum model
    and absorption line  features in the Cir X--1  first order HEG and
    MEG spectra for the three time-selected spectra during episodes of
    flaring    activity.      Line    flux    is     in    units    of
    photons/cm$^{2}$/s$^{-1}$.   Blackbody  luminosity  is  calculated
    considering a distance to the  source of 6 kpc.  Uncertainties are
    reported at  90\% confidence level  for a single  parameter; upper
    limits at 95\% confidence level.}  }
\label{tab3}
\end{deluxetable}
%%%%%%%%%%%%%%%%%%%%%%
%%%%%%%%%%%%%%%%%%%%%%%%%%%%%%%%%%%%%%%%%%%%%%%%%
\end{document}